\documentclass[german,english]{bvm2020}

\def\articlenumber{0000}

\date{}
\title{Multi-Channel Volumetric Neural Network for Knee Cartilage Segmentation in Cone-beam CT}

\titlerunning{Multi-Channel Volumetric Neural Network for Knee Cartilage Segmentation}

\author{Jennifer~Maier$^{1,2}$, Luis~Carlos~Rivera~Monroy$^1$, Christopher~Syben$^1$, Yejin~Jeon$^3$, Jang-Hwan~Choi$^3$, Mary~Elizabeth~Hall$^4$, Marc~Levenston$^4$, Garry~Gold$^4$, Rebecca~Fahrig$^5$, Andreas~Maier$^1$}

\authorrunning{Maier et al.}

\institute{%
$^1$Pattern Recognition Lab, Friedrich-Alexander-Universit\"at Erlangen-N\"urnberg (FAU), Erlangen, Germany\\
$^2$Machine Learning and Data Analytics Lab, Friedrich-Alexander-Universit\"at Erlangen-N\"urnberg (FAU), Erlangen, Germany\\
$^3$College of Engineering, Ewha Womans University, Seoul, Korea\\
$^4$Stanford University, Stanford, California, USA\\
$^5$Siemens Healthcare GmbH, Erlangen, Germany\\}

\email{jennifer.maier@fau.de}

\begin{document}

%
\selectlanguage{english}

\maketitle

\begin{abstract}
Analyzing knee cartilage thickness and strain under load can help to further the understanding of the effects of diseases like Osteoarthritis.
A precise segmentation of the cartilage is a necessary prerequisite for this analysis.
This segmentation task has mainly been addressed in Magnetic Resonance Imaging, and was rarely investigated on contrast-enhanced Computed Tomography, where contrast agent visualizes the border between femoral and tibial cartilage.
To overcome the main drawback of manual segmentation, namely its high time investment, we propose to use a 3D Convolutional Neural Network for this task.
The presented architecture consists of a V-Net with SeLu activation, and a Tversky loss function.
Due to the high imbalance between very few cartilage pixels and many background pixels, a high false positive rate is to be expected.
To reduce this rate, the two largest segmented point clouds are extracted using a connected component analysis, since they most likely represent the medial and lateral tibial cartilage surfaces.
The resulting segmentations are compared to manual segmentations, and achieve on average a recall of 0.69, which confirms the feasibility of this approach. 
\end{abstract}


\section{Introduction}
\begin{figure}[t]
	\setlength{\figbreite}{0.25\textwidth}
	\centering
	\subfigure[]{\includegraphics[height=0.3\textwidth]{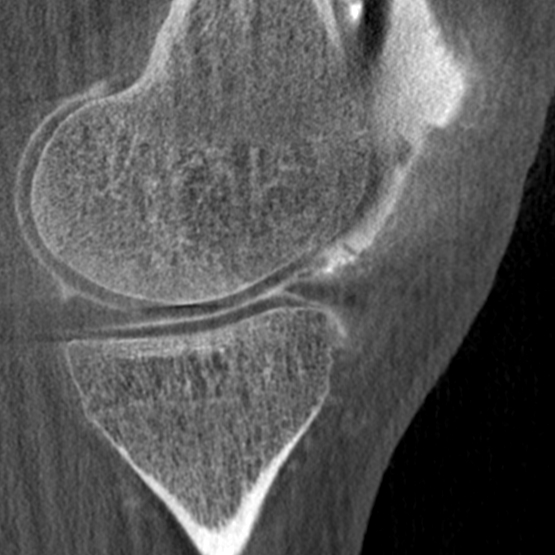}}
	\subfigure[]{\includegraphics[height=0.3\textwidth]{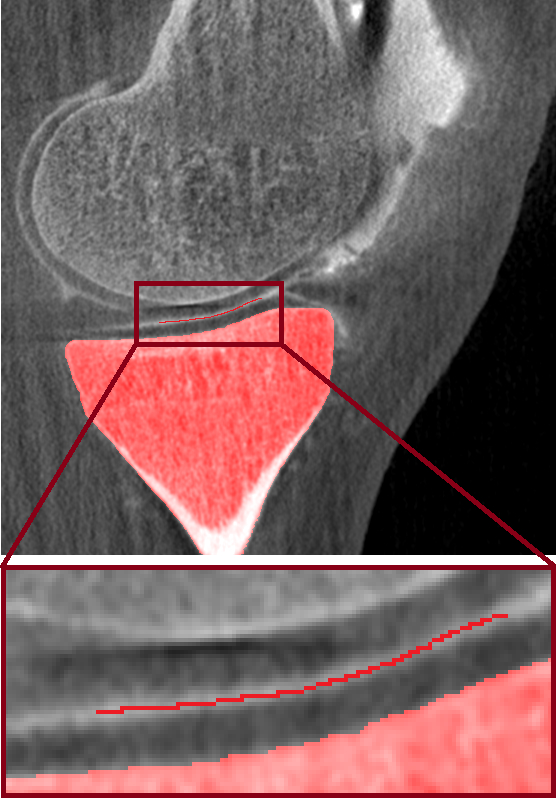}}
	\subfigure[]{\includegraphics[height=0.3\textwidth]{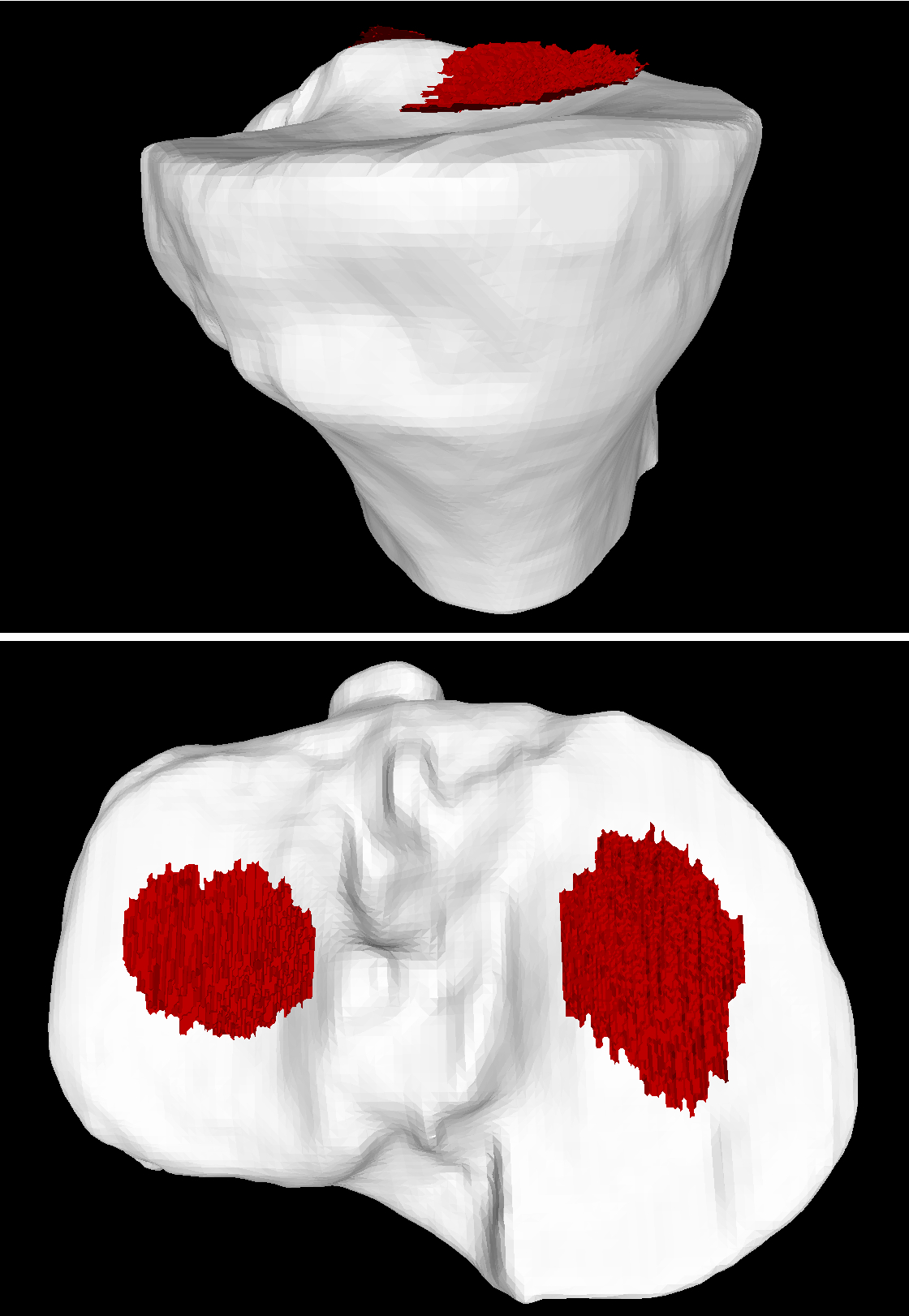}}
	\caption{Example volume and segmentation. (a)~Contrast-enhanced gray-scale image. (b)~Ground truth segmentation of bone and cartilage (red), zoomed image below. (c)~3D visualization of the segmentation in sagittal (top) and axial (bottom) view.}
	\label{0000-fig-01}
\end{figure}

Patients suffering from Osteoarthritis (OA) experience pain in their joints due to the degeneration of cartilage and bones. 
To better understand how OA is affecting the knee joint, it can be analyzed under load, because it then shows different mechanical properties compared to the unloaded case~\cite{0000-02}.
This analysis can be realized using weight-bearing \textit{in-vivo} cone-beam computed tomography (CBCT) acquisitions with injected contrast agent visualizing the thin line between femoral and tibial cartilage (Fig.~\ref{0000-fig-01}a).

A prerequisite for the analysis of cartilage is a prior segmentation of the knee's structures.
The segmentation of cartilage resp. its surface has mainly been investigated in Magnetic Resonance (MR) acquisitions~\cite{0000-04}. 
Since the conventional manual labeling of cartilage in CBCT is very time consuming, (semi-) automatic approaches using machine learning
have been developed.
Acetabular cartilage in the hip joint was segmented using a shape-based approach and prior knowledge~\cite{0000-05}, or by applying a seed-growing algorithm~\cite{0000-06}, both exploiting the specific shape of the hip joint.
Regarding the segmentation of the thin contrast agent line in knee CBCT, Myller et al.~\cite{0000-07} were one of the first to apply a semiautomatic approach based on model registration and intensity changes.
Their approach yielded good results on high resolution unloaded CT images segmenting the whole femoral and cartilage surface.

In contrast to this, this work aims to segment only the region of the contrast agent line where femoral and tibial cartilage are in contact.
Consequently, the main challenge is the high imbalance in the data between the small contrast agent line and the large background.
We propose an automatic segmentation based on a 3D volumetric convolutional neural network.
The network is trained and evaluated on manual segmentations of contrast enhanced knee CBCT volumes. 
Since the resulting segmentations contain many false positives as expected due to the high class imbalance, a post processing step of extracting the largest connected point clouds is applied.
%
%
\section{Materials and methods}
%

\subsection{Data}
The dataset used in this work was acquired under an IRB-approved protocol, containing in total 40 CBCT scans of 8 subjects in a supine (s) or weight-bearing (w) position.
The C-arm (Artis Zeego, Siemens Healthcare GmbH, Erlangen, Germany) acquired 496\,(s)/248\,(w)\,projections of size 1240$\times$960 pixels with isotropic pixel size of 0.308\,mm on a calibrated vertical\,(s)/horizontal\,(w) trajectory.
Contrast agent was injected in the knee to visualize the outline of soft tissues.
The reconstructions had a size of $512^3$\,voxels with an isotopic spacing of 0.2\,mm.

The tibia and the thin contrast agent line where tibial and femoral cartilage are in contact were manually segmented slice by slice in the sagittal view by an expert (Fig.~\ref{0000-fig-01}b, c).
In total, only 0.18\% of all voxels belonged to the cartilage surface (=~ positive voxels), resulting in a high imbalance in the annotations.

The dataset was divided into $70\%-10\%-20\%$ for the training/validation/test group, with no subject being represented in both training/validation and test.
Due to GPU memory restrictions, the data had to be sub-sampled into smaller volumetric patches of size $100^3$.
To address the high class imbalance, for training and validation the data was oversampled by using 70\% patches that have a randomly picked positive voxel in the center, and 30\% all negative patches. 
Four patches per volume were extracted for training and validation, and data augmentation was applied to the training patches with random rotations of $90^{\circ}$, $180^{\circ}$, and $270^{\circ}$.
For testing, the whole volumes were divided into disjoint patches.
%
%
\subsection{Multi-channel volumetric neural network}
\begin{figure}[tb]
    \centering
	\includegraphics[width=\textwidth]{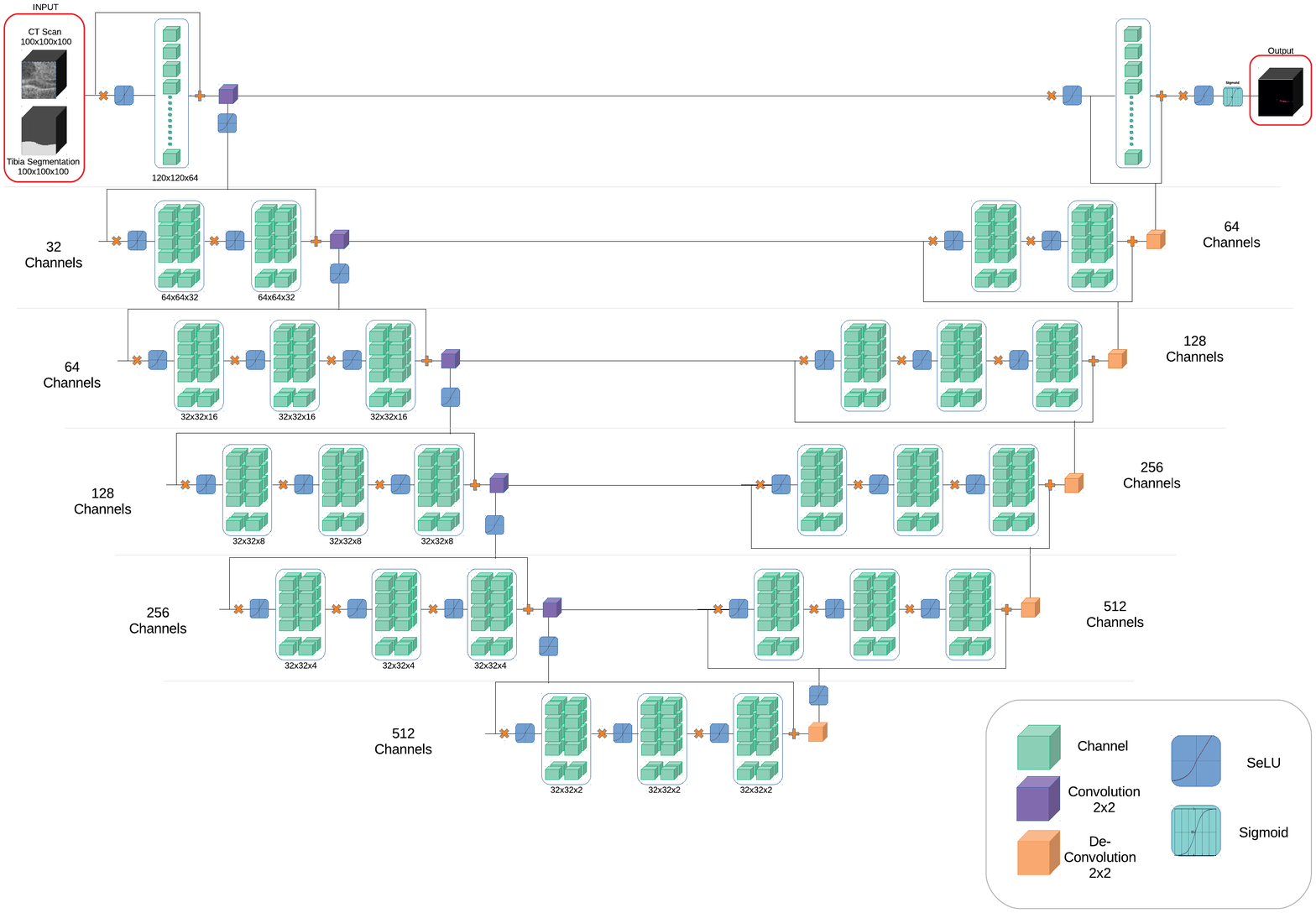}
	\caption{Proposed architecture.}
	\label{0000-fig-02}
\end{figure}
%
The architecture we used was a VNet~\cite{0000-08}, an extension of UNet
for volumetric data (Fig. \ref{0000-fig-02}).
It takes advantage of 3D convolutions, fully connected connections and modified types of residual connections.
Introducing skip connections between encoder and decoder path
produced results correctly located and at the same time with more confidence in the prediction.
The number of convolutions and several stages were adapted to the task of cartilage segmentation.
SeLu was used as the activation function showing a stable and relatively fast convergence.
Finally, AMSGrad was chosen as the optimization scheme, since it outperformed the ADAM traditionally used on VNet.
AMSGrad proved that adding the concept of memory for a highly imbalanced dataset produced better results and a faster convergence~\cite{0000-13}.
To avoid overfitting, dropout was added with a value of $60\%$.
%
%
\subsection{Loss function}
The Tversky index~\cite{0000-14} described by Equation~\ref{0000-eq1} was chosen as loss function since it is able to work with highly imbalanced data.
$p_{ni}, p_{pi}$ represent the negative and positive voxels of the prediction, $g_{ni}, g_{pi}$ are the negative and positive voxels in the ground truth annotation.
Tversky directly takes into account the relation between False Positive (FP) and False Negatives (FN) predictions and proposes parameters $\alpha$ and $\beta$ to manage the trade-off between both errors.
For this specific case, the highest performance was achieved using $\alpha=0.4$ and $\beta=0.6$.

\begin{equation}
    \label{0000-eq1}
    \begin{split}
    T(\alpha, \beta) = \frac{\sum_{i=1}^Np_{ni}g_{ni}}{\sum_{i=1}^Np_{ni}g_{ni}+\alpha\sum_{i=1}^Np_{ni}g_{pi}+\beta\sum_{i=1}^Np_{pi}g_{ni}}
    \end{split}
\end{equation}

%
%
%
\subsection{Connected component analysis}
To reduce the high number of false positive predictions due to data imbalance, the resulting segmentations were post-processed with a connected component analysis.
Since the two surfaces of medial and lateral cartilage are expected to be the largest segmented connected point clouds, all but the two largest connected components were discarded.
If this assumption didn't hold, a manual selection of the point clouds corresponding to the cartilage surface was performed.
%
%
\section{Results}
\begin{figure}[tb]
	\setlength{\figbreite}{0.25\textwidth}
	\centering
	\subfigure[]{\includegraphics[height=0.3\textwidth]{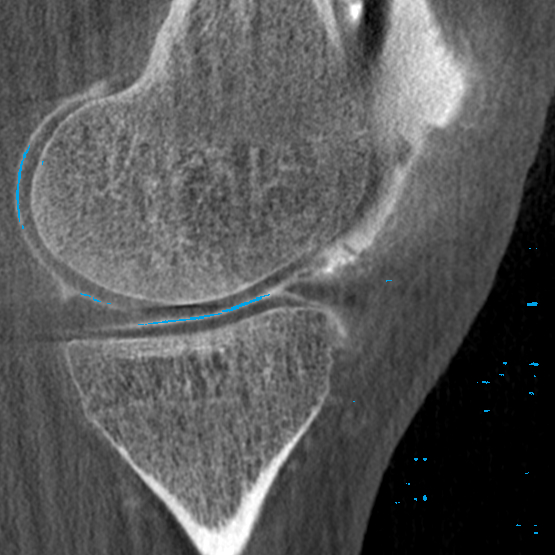}}
	\subfigure[]{\includegraphics[height=0.3\textwidth]{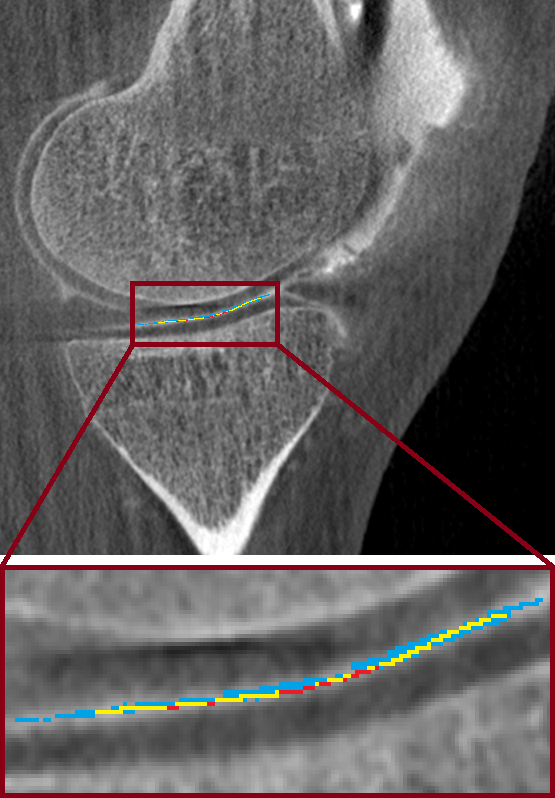}}
	\subfigure[]{\includegraphics[height=0.3\textwidth]{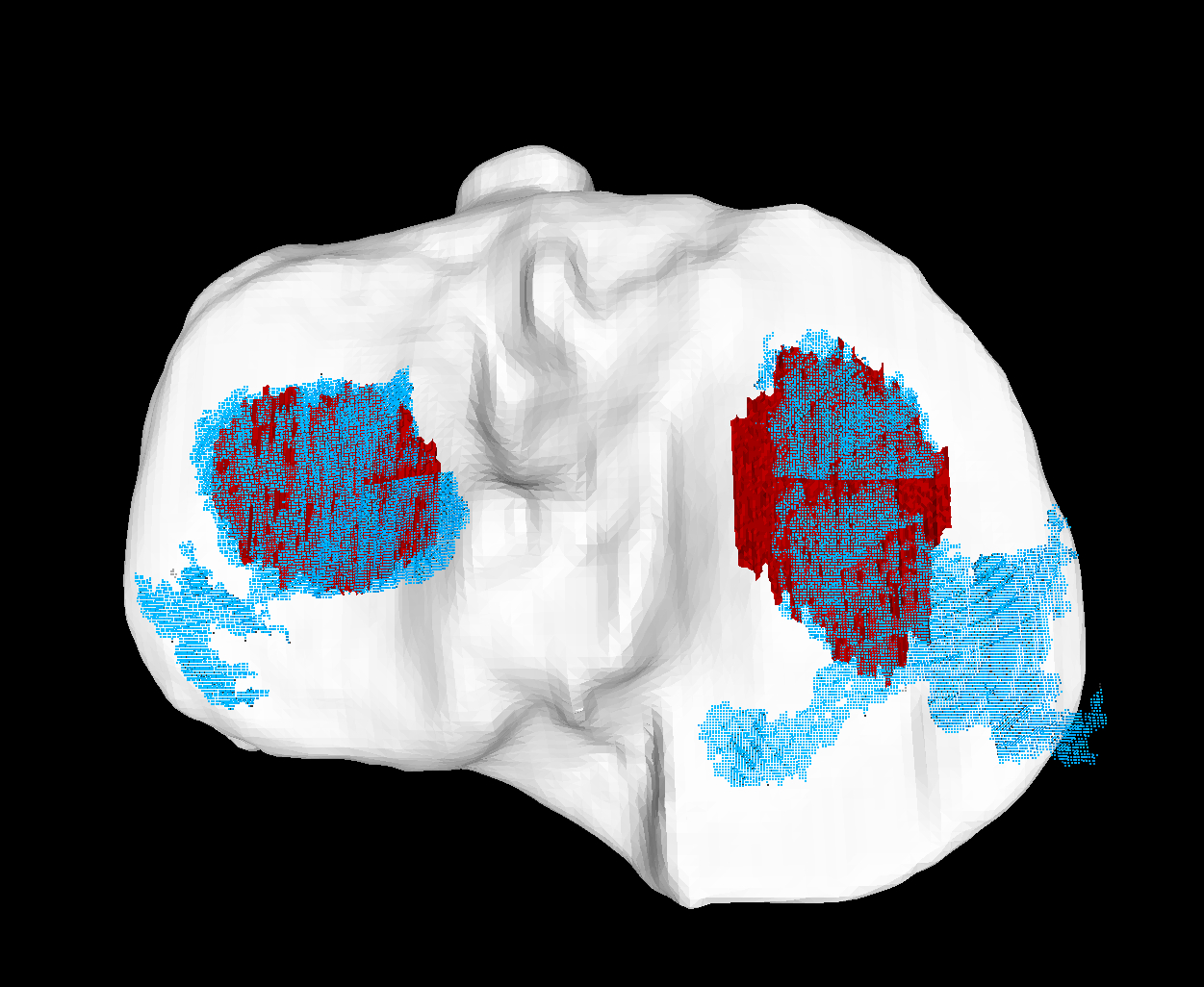}}
	\caption{(a)~Sagittal slice overlaid with network output (blue). (b)~Overlay of ground truth (red) and final output (blue) in a sagittal slice, overlap marked in yellow. (c)~3D view of ground truth (red) and final output (blue).}
	\label{0000-fig-03}
\end{figure}

To evaluate the network's performance the metrics accuracy, precision, recall, and dice index were computed.
An accuracy of 99\% was achieved due to the high number of negatives correctly classified.
An average recall of 0.69, precision of 0.24, and dice index of 0.35 were achieved.
Figure~\ref{0000-fig-03}a shows one slice of the network output containing many false positives. After the connected component analysis, the ground truth labels and the predictions show a high overlap (Fig.~\ref{0000-fig-03}b and c).
The connected component analysis successfully chose the correct patches in most of the test cases, and only one had to be adapted manually.
%
%
%
\section{Discussion}
The proposed network shows promising results for the task of knee cartilage surface segmentation.
Despite the use of oversampling and Tversky loss, the high imbalance still led to a high false positive rate.
Using mainly patches containing positive voxels for training led the model to learn that even patches in the periphery of the knee joint should contain positive voxels (Fig.~\ref{0000-fig-03}a).
Since these peripheral false segmentations are small and closely connected, the connected component analysis applied in post-processing was able to remove them and predict the desired cartilage segmentation in a stable way (Fig.~\ref{0000-fig-03}b).
Only the false positives in the segmentation's proximity could not be removed (Fig.~\ref{0000-fig-03}c).

We see the connected component post-processing step as an intermediate solution.
In future, we want to investigate an enhancement of the network using the prior knowledge about the segmentation being a 1D continuous line in the sagittal view.
This can be achieved following the learning with known operators paradigm~\cite{0000-15} by including either the connected component analysis or a polynomial fitting step directly into the network.

An additional reason for the high false positive rate is the current way of dividing the volume into patches, thereby restraining the network from learning the spatial relation of the cartilage contact area between femur and tibia.
The border between patches can even be seen in the resulting segmentation (Fig.~\ref{0000-fig-03}c).
The reason for dividing the volume into patches is the hardware limitation due to the large size of medical data.
A solution with bigger patches or even the full volume could be achieved using reversible networks as proposed in~\cite{0000-16}.

Note that the manual segmentations used as ground truth are one pixel thin lines in the sagittal view, meaning that a 1-pixel shift directly results in false predictions.
However, the contrast agent in the cartilage contact area is in most cases multiple pixels thick, leading the network to predict a point cloud instead of only a thin line (Fig.~\ref{0000-fig-03}b).
The consequence of this is directly observable in our reported metrics with a very low precision due to many false positives, but also with a good recall because most of the true labels are contained in the predicted point clouds.
As these metrics are used to compute the loss function and therefore guide the training, we hope that the enrichment of the network with prior knowledge or a polynomial fitting can stabilize the training and overcome this instability.

The presented results confirm the complexity of this highly imbalanced task, but show promising results towards a fully automatic cartilage segmentation in CBCT. Even though there are still many false positives in the final segmentation (Fig.~\ref{0000-fig-03}c), the proposed method can help to facilitate and accelerate the process of analyzing cartilage thickness in the clinical field.
%

\ack{This work was supported by the Research Training Group 1773 Heterogeneous Image Systems, funded by the German Research Foundation (DFG). Further, the authors acknowledge funding support from NIH 5R01AR065248-03 and NIH Shared Instrument Grant No. S10 RR026714 supporting the zeego@StanfordLab.}
%
%
\bibliographystyle{bvm2020}

\bibliography{0000}
\marginpar{\color{white}E\articlenumber} 
\end{document}